# New method for the study of psychotropic drug effects under simulated clinical conditions

Kudryavtseva N.N., Avgustinovich D.F., Bondar N.P., Tenditnik M.V., Kovalenko I.L., Koryakina L.A.

Neurogenetics of Social Behavior Sector, Institute of Cytology and Genetics SD RAS, Novosibirsk, 630090, Russia, e-mail: natnik@bionet.nsc.ru

**Abstract**

The sensory contact model allows forming different psychopathological states (anxious depression, catalepsy, social withdrawal, pathological aggression, hypersensitivity, cognition disturbances, anhedonia, alcoholism etc.) produced by repeated agonistic interactions in male mice and investigating the therapeutic and preventive properties of any drug as well as its efficiency under simulated clinical conditions. This approach can be useful for a better understanding of the drugs' action in different stages of disease development in individuals. It is suggested that this pharmacological approach may be applied for the screening of different novel psychotropic drugs.

## Introduction

Screening of newly synthesized drugs (agents) with putative therapeutic properties includes the stages of preclinical and clinical trials. In the stage of preclinical trials, tests on animals are performed to determine such general drug characteristics as lethality, toxicity, absorption and excretion rates and also to study the mechanisms of action. Then new drugs are compared with a well-known medical analog using the experimental methods adopted in International pharmacopoeia. If positive properties of the drug intended for the treatment of a disease are confirmed, the drug has to undergo clinical trials when its action is tested in a group of patients. Such studies are usually designed so that patients who agreed to participate in the experiment are given placebo or the drug tested. A group of patients provided with conventional treatment or a group of people with the same disease untreated for some reasons is used as control.

The placebo method, however, is traditionally blamed from the moral and ethical viewpoints as in the variants with both placebo and a novel drug a patient becomes the object of experimentation and is likely to be deprived of necessary treatment or provided with treatment not so effective as it could have been otherwise. It sometimes happens that volunteers become the victims of side effects of the novel drug, which had not been revealed in animals. One of common difficulties of clinical trials is the selection of volunteers for the experiments among people suffering a given disease. In any case, novel drugs have to be studied on patients because the reaction of a diseased organism to the administration of the drug may differ drastically from that of a healthy one.

The suggested approach, in our view, makes it possible if not to do totally without clinical trials but to minimize them at least for psychotropic drugs used to correct different psychoemotional disorders and accompanying psychosomatic diseases.

Experimental approach is based on the model of chronic social conflict or the so-called sensory contact model [62], which was originally developed to study the mechanisms of aggressive and submissive behaviors of male mice. It was shown that repeated experience of social victories or defeats in daily agonistic interactions leads to the formation of persistent opposing kinds of social behavior in male mice – the winners (aggressors) and losers (victims of aggression). Depending on emotional state (positive or negative) of an individual, multiple neurochemical alterations in the synthesis, catabolism and receptors of key brain neurotransmitters are followed by behavioral and physiological changes. In general, it was shown that repeated experience of aggression is accompanied by the activation of the dopaminergic and opioidergic systems and inhibition of serotonergic system of the brain [49, 56] while repeated experience of social defeats leads to the attenuation of the activity of all three mediator systems [8]. As a consequence, the winners and losers were found to exhibit significant differences in emotionality, motor and exploratory activities, level of sociability, alcohol intake as well as in the state of immune system and gonadal function [67]. It was also shown that long exposure to social confrontations leads to the formation of psychoemotional and somatic disturbances [8, 49, 59, 60, 64] with the forming behavioral pathology depending on both the social behaviors and the duration of agonistic interactions. Besides, in mice with heredity-specific features of neurochemical regulation and psychoemotional characteristics (different inbred strains), repeated experience of social interactions generated the changes differing in the degree of manifestation and sometimes in nature per se [50]. The model could be used in medical-biological and basic studies [40] and also for the study of a broad spectrum of problems of social biology and biological psychiatry.

Therefore, on the one hand, the sensory contact model is useful in generating and investigating different psychoemotional and psychosomatic disturbances in animals. On the other hand, it gives the opportunity of using animals with behavioral pathology to investigate the action of novel (along with widely used) psychotropic drugs and conduct their screening in the simulated clinical conditions. In this respect, it would be useful to outline possible applications of the proposed experimental method for detecting therapeutic and protective effects and efficacy of prospective psychotropic drugs. We are very grateful to I.P.Ashmarin, who was first to notice and drew our attention to possible unique application of the model in pharmacological screening of drugs and encouraged us to further elaborate this problem.

## Psychoemotional and psychosomatic pathologies generated in male mice under sensory contact model

The sensory contact model has been described in detail many times earlier [8, 59, 62]. Pair of males lives in common cage devided into half by transparent perforated partition. Key principle is in the formation of opposing patterns (aggressive or submissive) of social behaviors in male mice by subjecting them to daily agonistic interactions. A long experience of social confrontations has led to changes in behavior of the winners and losers in various situations. In our studies, eight criteria used were thought to point to the formation of behavioral pathology [49]:

- Change (increase or decrease) in the **duration** and/or **expression** of demonstration of behavioral forms;
- Emergence of novel behavioral form, which have not been demonstrated by animals before;
- **Inadequacy** of behavioral response or physiological reaction to some social or environmental stimuli; uncontrollable behavior in some circumstances;
- **Inadaptability** of behavior in given environmental conditions or experimental situations;
- **Generalization** of dominating motivation;
- Prolonged **retention** (persistency) of changes in behavior and psychoemotional state after the cessation of psychopathogenic factor action;
- Expressed multiple **neurochemical alterations** in the brain;
- **Similarity** of behavioral pathology occurring in mice to clinical picture of a human disease: similarity of etiology, symptomatology, sensitivity to the drugs used for the treatment of such a disease and also the similarity of neurochemical alterations arising in mice and humans as the disease progresses.

For example, it was demonstrated that in males of the C57BL/6J (C57) strain a long experience of social defeats has led to the development of mixed anxiety-depression state meeting all formal criteria suggested [30, 79] for the experimental model of depression, which is similar to the clinical manifestation of the disease in humans [8, 59]. At the same time, the same stress of social confrontations in defeated males of the CBA/Lac (CBA) strain has led to the formation of pronounced catalepsy demonstrated by mice in free behavior and in behavioral tests [72, 75]. Repeated experience of aggression reinforced by victories resulted in the development of pathological aggression accompanied by signs of mania, hyperactivity, addictions etc. [49]. **Table 1** summarizes references to the studies that showed the formation of various behavioral pathologies in male mice of both strains under repeated experience of social victories or defeats. The most extensive studies have been conducted and the most satisfactory validating results obtained on mice for **anxious depression, generalized anxiety, pathological aggression** and **psychogenic immune deficiency.**

## Investigation of therapeutic (medicative) effect of drugs with prospective psychotropic properties

The general design of the experiments is as follows: during 20-30 days a psychoemotional disorder is induced in male mice by repeated social agonistic interactions, which are inevitably followed by psychosomatic changes. Then the defeated animals are put into comfortable housing conditions (see the *Footnote*[*]) to start chronic (1-2 times a day for at least 2 weeks) administration of the drug under investi-gation. At the same time, other group of "sick" animals is treated with placebo (vehicle) in analogous conditions (**Fig. 1**). Upon the completion of "treatment" the two groups of animals are compared (using relevant behavioral tests, see below) to detect therapeutic effect of the drug. By comparing

---

[*] Comfortable housing of animals after the development of behavioral psychopathology implies the improvement of housing conditions, which, however, may differ by the degree of aversion of the situation. One variant of comfortable habitation is to keep an animal with a partner separated by a partition but without daily confrontations. In this variant the physical element of impact is removed without social isolation while the psychoemotional influence remains. In the second variant male is put in a cage with females. In this variant physical element and permanent fear to be attacked by an aggressive partner are absent and fear motivation is replaced with sexual motivation.

**Table 1.** Different psychoemotional and somatic pathologies developing in aggressive and submissive male mice under chronic social confrontations

| SUBMISSIVE MICE | | | AGGRESSIVE MICE | | |
|---|---|---|---|---|---|
| Behavioral Pathologies and Psychosomatic Disorders | Strain | References | Behavioral Pathologies and Psychosomatic Disorders | Strain | References |
| Mixed anxiety/depression state Generalized anxiety | C57 | [8, 11, 12, 26, 59, 60, 64] | Pronounced anxiety | CBA C57 | [12, 21, 54] |
| Behavioral deficit | C57 CBA | [59, 60, 64] | Hyperactivity Maniac state (?) | CBA | [49, 68] |
| Cognition dysfunction | C57 | [34, 35, 39] | Cognition dysfunction | C57 | [20] |
| Increased ethanol intake | C57 | [51, 71] | Addiction to aggression | C57 | [49] |
| Catalepsy | CÀÀ | [63, 72,73, 75] | Stereotypic behavior | CBA | [56, 75] |
| Low communication Social withdrawal (autism?) | C57 CBA | [12, 16, 59-61, 64] | Pathological aggression Enhanced impulsivity | CBA C57 | [23, 49, 68] |
| Decreased pain sensitivity | C57 | [8] | Hypersensitivity, Enhanced irritability | C57 CBA | [23, 49] |
| Sexual dysfunction Decreased fertility | C57 | [5 , 42, 60] | Sexual dysfunction | CBA C57 | [5, 52, 68] |
| Psychogenic immune deficiency Increased metastasis | CBA C57 | [3, 33, 37, 42, 43, 84, 91] | Learned aggression | CBA C57 | [21, 49, 56, 58] |

"sick" animals, which were chronically treated with placebo, with drug-treated animals it is possible to detect positive or negative effect of the drug. Along with experimental animals the behavior of intact animals (control, norm), which have not been exposed to any treatment, is studied (see *Footnote***[†]). By comparing the control with "sick" individuals, who have been treated with the drug, the efficiency of the drug action could be determined, i.e. to what extent the drug has improved the behavioral or physiological characteristics as compared with those in the control (intact animals).

The behavior of animals is studied in such commonly used tests as "open-field" to measure motor and exploratory activity and the level of anxiety and emotionality; "elevated plus-maze", which is sensitive to the anxiolytics, to evaluate the level of anxiety in animals [77, 83, 88]; "exploratory activity" to estimate animal behavior in new conditions, Porsolt test, which is sensitive to the action of antidepressants [32, 85], to determine the level of depression by the intensity of helplessness behavior in unavoidable stressing situation; "hot-plate test" to evaluate sensitivity to pain and other tests. In our studies, "partition test" is also commonly used [55, 61] to estimate the behavioral reaction of male mice to the partner – communicative behavior

(sociability) that may include different motivations depending on the design and purposes of the experiment. In particular, in aggressive males this test is used to assess aggressive motivation [54], in submissive males – anxiety, social withdrawal [12]. Besides, social behavior of two partners during agonistic interaction is studied using V.P. Poshivalov's behavioral atlas [86] for the identification of respective behavioral patterns in mice.

The mixed anxiety/depression state in mice [8, 59] was used to study the properties of three drugs (**Table 2, Fig. 2**). These are fluoxetine, the drug widely used for treatment of depression in humans [87, 90], novel drugs fluoglyzine [92], analog of the fluoxetine, and noolit [24] (lithium-based enterosorbent). Fluoxetine (as fluoglyzine) was shown to have antidepressant (in Porsolt test) and anxiolytic effects, enhancing sociability in the "partition" test [7]. Fluoxetine also enhanced exploratory activity. Noolit was shown to produce pronounced antidepressant and anxiolytic effects [25, 26].

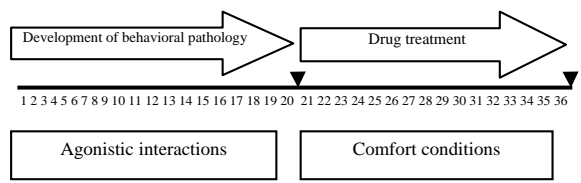

**Fig.1.** Regimen of medicative treatment. See the text for explanation.

---

** As control intact animals are used, who are housed individually for 5 days, which is thought to remove the effect of group interactions without producing the effect of social isolation [62]. It is thought to be the best control of all possible controls for given conditions [6].

**Table 2**. Effects of chronic preventive and medicative drug treatment on submissive C57 male mice.

| Drugs | Dose, treatment | Behavioral tests | Effects | References |
|---|---|---|---|---|
| **Tianeptine,** selective serotonin reuptake enhancer | **Preventive treatment** 10 mg/kg/day, i.p., 14 days | Partition test Open field test Exploration test Porsolt test | No effect Weak anxiolytic No effect Antidepressive | [60] |
| **Imipramine,** tricyclic antidepressant, serotonin and norepinephrine reuptake inhibitor | **Preventive treatment** 10 mg/kg x 2 times/day, i.p., 14 days | Partition test Open field test Exploration test Porsolt test Pain sensitivity | Anxiogenic Weak anxiolytic No effect Antidepressive Nociceptive | [64] |
| **Fluoxetine,** selective serotonin reuptake inhibitor | **Medicative treatment** 15 mg/kg/day, p.o., 14 days | Partition test Plus-maze test Open field test Porsolt test | Anxiolytic No effect Increased exploration Weak antidepressive | [7] |
| | **Preventive treatment** 25 mg/kg/day, p.o., 14 days | Partition test Plus-maze test Open field test Porsolt test | No effect No effect No effect Prodepressive | [48] |
| **Citalopram,** selective serotonin reuptake inhibitor | **Preventive treatment** 10 mg/kg x 2 times/day, 14 days | Partition test Open field test Porsolt test Pain sensitivity Exploration test | No effect No effect No effect No effect No effect | [70] |
| **Buspirone,** 5-HT$_{1A}$ receptor agonist | **Preventive treatment** 1 mg/kg/day, i.p., 14 days | Partition test Plus-maze test Porsolt test | No effect Anxiolytic Weak antidepressive | [9] |
| **Ipsapirone,** 5-HT$_{1A}$ receptor agonist | **Preventive treatment** 3 mg/kg/day, i.p., 14 days | Partition test Plus-maze test Porsolt test | No effect Anxiolytic No effect | [9] |
| **Novel drugs with putative antidepressant and anxiolytic effects** | | | | |
| **Noolit,** lithium-containing enterosorbent [15] | **Medicative treatment** 665 mg/kg, p.o., 18 days | Partition test Plus-maze test Open field test Porsolt test | No effect Anxiolytic Anxiolytic Antidepressive | [25, 26] |
| | **Preventive treatment** 665 mg/kg, p.o., 18 days | Partition test Plus-maze test Open field test Porsolt test | No effect Anxiolytic No effect Antidepressive | |
| **Fluoglyzine,** analogue of fluoxetine [41] | **Medicative treatment** 15 mg/kg/day, p.o., 14 days | Partition test Plus-maze test Open field test Porsolt test | Anxiolytic No effect No effect Antidepressive | [7, 48] |
| | **Preventive treatment** 25 mg/kg/day, p.o., 14 days | Partition test Plus-maze test Open field test Porsolt test | No effect No effect No effect Prodepressive | |

*Note:* * Weak anxiolytic or weak antidepressive effect means tendency (0.05<p<0.1).

It was assumed that the regimen of therapeutic treatment (**Fig. 1**) imitates the conditions of clinical treatment of patients. It should be noted that in the experiments positive action of the drugs was manifested after two weeks of treatment. However to verify a sustained effect of the drugs they are to be administered for a longer time, possibly, two months, similarly to the duration of clinical treatment of depression in humans. Besides, additional studies of animals after treatment are needed to predict the probability and incidence of relapses, which are quite common in patients with psychoemotional disorders.

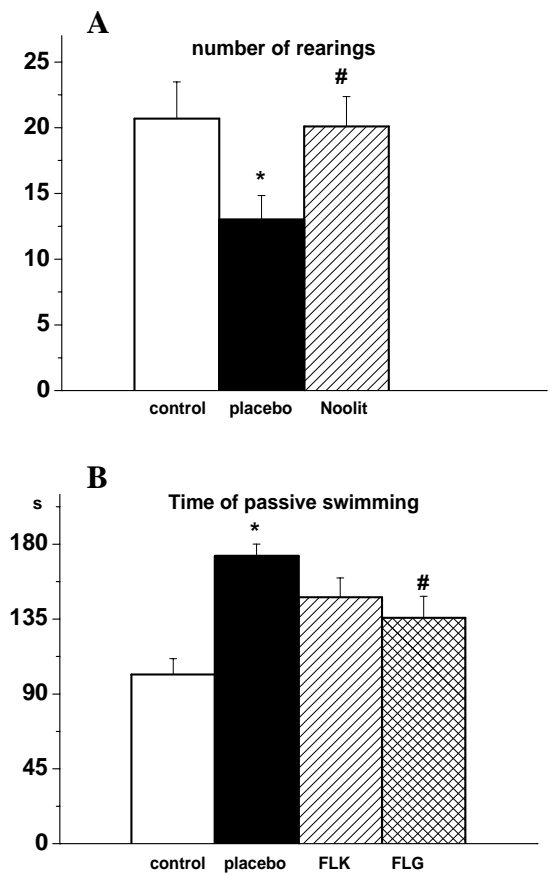

**Detection of protective properties of drugs**

As a rule, an individual is incapable of avoiding or at least minimizing negative influence of social environment and surroundings, in which he or she lives or has to stay in particular living circumstances. In such cases the questions arise on how to prevent the development of a disease under long exposure to psychopathogenic factors? Our approach allows detecting protective effects of drugs administered for preventive purposes under exposure to chronic agonistic interactions. For this purpose, after five days of social interactions forming opposite social behaviors, animals are treated chronically by drugs with assumed or known therapeutic properties

(**Fig. 3**). At the same time, in analogous way a group of animals receives a placebo. After a period of time, which is to be not shorter than two weeks for drugs with assumed psychotropic properties, all animals are investigated in behavioral tests. This regimen was used to administer the drugs (**Table 2, Fig. 4**) that are used in the clinical practice for reduction of depression and/or anxiety and that influence the serotonergic system, which undergoes alterations in the process of depression formation [8, 10, 70].

Buspirone, insapirone, tianeptine and noolit (but not fluoxetine, fluoglyzine, citalopram) produced anxiolytic effect, i.e. the level of developing anxiety on the background of chronic drug administration was lower at least in one of the behavioral tests as compared with placebo-treated animals, which points to protective effect of these drugs. Imipramine, tianeptine and noolit were shown to prevent the development of high level of depressiveness while buspirone appears to have a lower antidepressive potency. Citalopram and insapirone had no effect on depression in animals while fluoxetine and fluoglyzine produced even a pro-depression effect when administered under the described regimen.

Naturally, a question arises: what underlies different efficacy of medications? It could depend on the dosage and regimen of treatment. For example, in our experiments, citalopram in the dose of 10 mg/kg was not effective while in rats the same drug in the dose of 30 mg/kg when administered for a longer time prevented the development of depression [89]. It should not be excluded that under exposure to persisting powerful stress a complex psychotropic medication rather than mono-therapy with a single antidepressant is needed. Nevertheless, twenty years of work with the model have given us the understanding that different stages of the disease may require different treatments depending on the state of the brain neurotransmitter systems involved in the pathological process. It has been shown that state of the neurotransmitter activity undergoes changes during the development of psychoemotional disorders (see below).

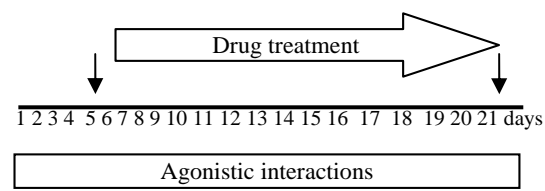

**Fig.3.** Regimen of preventive treatment. See the text for explanation.

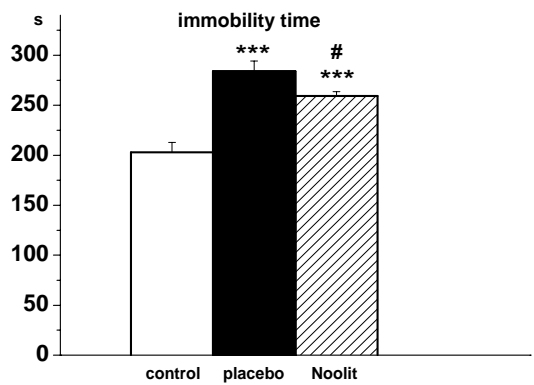

**Fig.4.** Effects of noolit (665 mg/kg) and placebo on the behavior of male mice in Porsolt's test under preventive treatment [26]. Noolit prevents the development of pronounced depression: animals receiving the drug and placebo significantly differ in the immobility time.

*Note*: *** p<0.01 - as compared with control; # p<0.05 - as compared with placebo

### Study of the drug effects depending on the stage of psychoemotional disorder

Neurochemical studies have produced a bulk of evidence that under repeated experience of aggression or social defeats the neurotransmitter systems undergo specific dynamic changes in synthesis, catabolism and receptors in male mice [8, 49, 60]. The consequences of chronic confrontations seem to be accumulated and the occurring neurochemical changes may differ depending on the duration of psychoemotional stress and the depth of developing behavioral pathology.

In our studies we develop pharmacological approach, which allows estimating possible changes in brain neurochemical activity during the formation of psychoemotional disorders. For this purpose, preliminarily chosen dose of a drug is administered into intact animals and animals in different stages of "disease". Comparison of the drug effect is conducted in the every experimental group by comparing the tested behavior of drug-treated animals with that of the vehicle-treated. Diverse effects of the drug as regards the intensity or direction in intact and "sick" mice point to a changed state of the mediator system involved in the pathogenic process. With the help of this method it was shown that different stages of disease are often characterized by dissimilar reaction to many drugs.

*Table 3* summarizes experimental data on the action of some drugs in submissive mice during the formation of mixed anxiety-depression state. In general, it was shown that in the control and in defeated individuals after 10 days of confrontations (T10 losers), 5-HT$_{1A}$ receptors agonist 8-OH-DPAT, selective agonist of D$_2$ receptors sulpiride and D$_1$/D$_2$ receptors antagonist cis-flupenthixol produced similar unidirectional effect, although of varying intensity [1, 2]. In animals after 20 days of confrontations (T20 losers) leading to apparent behavioral pathology no effect was observed, which could be the indication of underdoses resulting from the changes in the neurotransmitter activity. One exception was SKF 38393, the agonist of D$_1$ receptors, which produced a unidirectional activating (anxiolytic) effect on the behavior of defeated mice in the partition test, although the effect was more pronounced in T20 losers [1]. No effect was found in intact animals. In male mice with a short experience of social defeats (3 days) the opioid receptor inhibitor naltrexone in a low dose (0.25 mg/kg) caused the stimulation of active defense during the intermale confrontations and in a high dose (1 mg/kg) – its reduction. Naltrexone was totally ineffective in defeated males exposed to confrontations for 10 days [57].

Evidence in support of the fact that the efficacy of drugs depends on the stage of behavioral pathology was also found in experiments with aggressive mice (*Table 4*). For instance, haloperidol, a neuroleptic used in clinical practice for the treatment of acute psychotic states, was also effective in aggression reduction in male mice in the first confrontations and virtually ineffective in males with a long experience of aggression (T20 winners) [66]. Analogous effect was produced by SCH-23390, the antagonist of D$_1$ receptors [23]. Naltrexone decreased aggression (in a dose of 1 mg/kg) in CBA mice with a short experience of aggression (3 days) and was of no value in the winners with 10-day experience of confrontations (T10 winners) [57] or even produced a stimulating effect in C57 T20 winners [76]. Anxiolytics buspirone and diazepam (*Table 4, Fig. 5*) reduced aggression, exerting anxiogenic effect in T3 winners. In T20 winners diazepam produced anxiolytic effect and reduced aggression (*Table 4, Fig. 5*) while buspirone was ineffective [21, 22, 65]

An obvious implication of all the experiments is that psychoemotional disorders like many other ailments do not emerge out of the blue (all at once), this is rather a process in time, when the accumulating changes aggravate the state. Direct evidence was obtained in the study of the mouse brain serotonergic system during the formation of anxious depression, revealing dynamic alterations of serotonin synthesis, catabolism and receptors [8, 10, 70]. On the base of these studies it was concluded [8] that in animals during the first days of confrontations social stress evokes the activation of the serotonergic system. Subsequently, under systematic psychoemotional negative impact a hypofunction of the serotonergic system is formed at least in its limbic

**Table 3.** Effects of acute drug treatment on intact animals and submissive mice with 3 (T3), 10 (T10) and 20 (T20) days of social defeats in agonistic interactions.

| Drugs | Dose, treatment | Strain | Groups | Behavioral tests | Effects | References |
|-------|-----------------|--------|--------|------------------|---------|------------|
| **8-OH-DPAT,** 5-HT$_{1A}$ receptor agonist | 0.1, 0.5 mg/kg, s.c. 20 min 35 min 40 min | C57 | Intact | Partition test* Plus-maze test Porsolt test | Anxiogenic Anxiogenic (0.5 mg/kg) No effect | [+] |
| | | | T10 | Partition test Plus-maze test Porsolt test | No effect Anxiogenic (0.5 mg/kg) No effect | |
| | | | T20 | Partition test Plus-maze test Porsolt test | No effect No effect No effect | |
| **Sulpiride,** D$_2$ receptor antagonist | 20 mg/kg, i.p., 15 min | C57 | Intact | Partition test Porsolt test | Anxiogenic Prodepressive | [1, 2] |
| | | | T10 | Partition test Porsolt test | Anxiogenic Prodepressive | |
| | | | T20 | Partition test Porsolt test | No effect No effect | |
| **Cis-flupentixol,** D$_1$/D$_2$ receptor antagonist | 0.2 mg/kg, i.p., 20 min | C57 | Intact | Partition test Porsolt test | No effect Prodepressive | [1, 2] |
| | | | T10 | Partition test Porsolt test | No effect Prodepressive | |
| | | | T20 | Partition test Porsolt test | No effect No effect | |
| **SKF 38393,** D$_1$ receptor agonist | 10 mg/kg, i.p., 15 min | C57 | Intact | Partition test Porsolt test | No effect No effect | [1] |
| | | | T10 | Partition test Porsolt test | Anxiolytic No effect | |
| | | | T20 | Partition test Porsolt test | Anxiolytic Antidepressive | |
| **Naltrexone,** opioid receptors antagonist | 0.25, 1 mg/kg, i.p., 15 min | CBA | T3 | Active defense Active defense | Stimulation (0.25mg/kg) Decrease (1 mg/kg) | [57] |
| | | | T10 | Active defense Active defense | No effect No effect | |

*Note:* Here and in other tables: (0.5 mg/kg) – effective dose of drug was shown in brackets. Active defense was estimated during intermale interaction. * - partition test measures anxiety level in submissive mice. + - in press.

brain areas. A hypofunction has also developed in the dopaminergic brain systems.

Understanding the dynamic nature of neurochemical changes occurring in the brain implies differentiated approach to the therapy of a developing pathology, for example, depression, depending on its stage. If bad mood is accompanied by complex somatic disturbances (in particular, body weight changes, sexual dysfunction, suppression of immune system), as well as psychomotor retardation, immobility and apathy (indifference), which are characteristic of severe depression, one may suggest a hypofunction of the serotonergic systems and administer the drugs enhancing neurotransmission, assuming that such drugs could be of more effective as they should produce immediate positive effect. Obviously, the same drugs could aggravate the state if taken on the background of enhanced serotonergic activity, which is observable in the initial stages of the disease. Theoretically it can be predicted that in this stage drugs temporarily blocking the serotonergic transmission or arresting the state of anxiety and stress would be effective, which, in turn, will mitigate the influence of psychopathogenic factors inducing the development of depression.

We have obtained experimental evidence supporting these views (see *Table 2*). For example, fluoxetine and citalopram, selective blockers of serotonin reuptake, were ineffective on preventive administration under activated serotonergic activity observable in animals in the initial stage of depressive pathology. Imipramine, a nonselective reuptake inhibitor, even produced a weak anxiogenic effect in one of the tests in these conditions and

***Table 4.*** Effects of acute drug injection on intact animals and aggressive mice with 1 (T1), 3 (T3), 10 (T10) and 20 (T20) days of social victories in agonistic interactions.

| Drugs | Dose, Treatment | Strain | Groups | Behavioral tests | Effects | References |
|---|---|---|---|---|---|---|
| **Haloperidol**, $D_1/D_2$ receptor antagonist | 0.1, 0.4 mg/kg, i.p., 30 min | C57 | T1 | Partition test* Aggression | Decrease Decrease | [66] |
| | | | T20 | Partition test Aggression | No effect No effect | |
| **SCH-23390**, $D_1$ receptor antagonist | 0.1 mg/kg, i.p., 30 min | C57 | T1 | Partition test Aggression | Decrease Decrease | [23] |
| | | | T20 | Partition test Aggression | No effect No effect | |
| **Naltrexone**, Opioid receptors antagonist | 0.25, 1 mg/kg, i.p., 15 min | CBA | T3 | Aggression | Decrease | [57] |
| | | | T10 | Aggression | No effect | |
| | 0.25 mg/kg i.p., 15 min 1 mg/kg, i.p., 15 min | C57 | T1 | Partition test Aggression | Decrease Decrease | [76] |
| | 0.25, 1mg/kg, i.p., 15 min | | T20 | Partition test Aggression | No effect Increase | |
| **Diazepam** benzodiazepine derivative | 0.5 mg/kg, i.p., 2.5 hour | C57 | T3 | Aggression Plus-maze test | Decrease Anxiogenic | [65] |
| | | | T20 | Aggression Plus-maze test | Decrease Anxiolytic | |
| **Buspirone**, $5\text{-HT}_{1A}$ receptor agonist | 1 mg/kg, i.p., 30 min | C57 | Intact | Partition test Plus-maze test | No effect No effect | [22] |
| | | | T3 | Partition test Plus-maze test Aggression | Decrease Anxiogenic Decrease | |
| | | | T20. | Partition test Plus-maze test Aggression | No effect No effect No effect | |

*Note*: * - partition test measures the level of aggressive motivation in the winners [55].

simultaneously had an antidepressant effect, possibly due to its action on the catecholaminergic systems. At the same time fluoxetine, when administered therapeutically on the background of a hypofunction of the serotonergic activity was found to produce a positive effect.

Therefore, it is obvious that different stages of a disease require different drug correction depending on the status of brain neurotransmitter activity. In this respect the sensory contact model in mice, which allows monitoring the changes in neurochemical activity of the brain as a disease progresses from norm to deep pathology, could be used for screening psychotropic drugs to determine their efficacy in different stages of the pathological process. The knowledge of neurochemical changes taking place as psychopathology evolves would help in finding rational treatment and predicting the effects of novel drugs.

**Experimental approach to the personal therapy of psychoemotional disorders**

It is well known that the majority of modern drugs are effective in less than half of all patients and a number of strong medicines, in particular those used in oncology – in one third of all patients at best. This is thought to be due to individual peculiarities of diseased organisms having varying hereditary sensitivity to drugs. The role of hereditary factor can be elucidated in the experiments with animals of inbred strains reacting differently to the drugs administered. However, our experiments demonstrated that one and the same behavioral or physiological parameter under the influence of a drug could change in different ways in animals of one inbred strain but with the opposite patterns of social behavior – in the winners and losers. For example, in joint Dutch-Russian studies it was shown that U-50,488H, the agonist of *k*-opioid receptors (***Table 5***) [53] produced social withdrawal in the partition

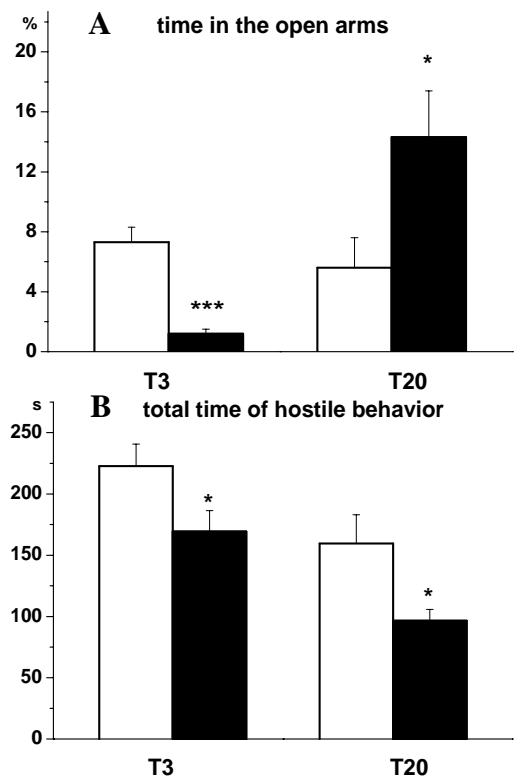

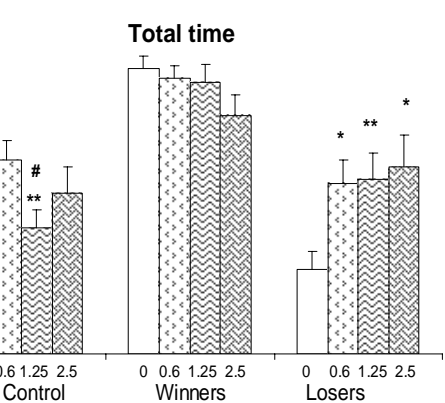

**Fig.6.** Effect of selective *k*-opioid receptor agonist U-50,488H (0.6, 1.25 and 2.5 mg/kg) on communicative behavior of the control males as well as of the winners and losers and after 10 days of social confrontations as estimated in the partition test [53].
*Note*: white columns – vehicle-treated animals; * p<0.05; ** p<0.01 – as compared with vehicle; # p<0.05 – as compared with previous dose of the drug.

**Fig.5.** Effect of single diazepam injection (0.5 mg/kg) in the winners after 3 (T3) and 20 (T20) intermale confrontations on the time spent in open arms of the plus-maze test (A) and on the time of hostile behavior toward the partner (B) [65]. Diazepam produced anxiogenic effect in males with a short experience of aggression (T3) and anxiolytic effect in males with a long experience of aggression (T20) and reduced the total time of hostile behavior (attacks, threats etc) in both groups of males.
*Note*: white columns – vehicle-treated animals; black columns – diazepam-treated animals. * p<0.05; *** p<0.001 – as compared with vehicle.

(**Fig. 6**). At the same time, DAGO (DAMGO), the *µ*-opioid receptor agonist produced social withdrawal in the partition test and anxiogenic effect in the plus-maze test in the control and winners and had no effect in the losers [53]. It is worth noting that animals of different social groups incipiently demonstrate significant behavioral differences in this test. Chronic administration of U-50,488H did not significantly affect the intake of ethanol under the free two bottles choice paradigm (water, ethanol) in the winners and increased it in the losers (**Fig. 7**). The dopamine D₂ receptors agonist (+)3PPP had no effect on conditioned passive avoidance reaction (CPAR) in intact animals, inhibited it in the winners and enhanced in the losers [39] (**Table 5**).

Different reaction to the administration of the same drug was also found in the studies of physiological response in animals with different social status. In particular, acute administration of serotonin into the 3d ventricle of the brain produced test in control animals, had no effect on the winners and produced a marked anxiolytic effect on the losers different changes in the gastric mucosa (number of hemorrhages and erosions) of the winners and losers with 3 days of confrontations [47] (*Fig. 8*).

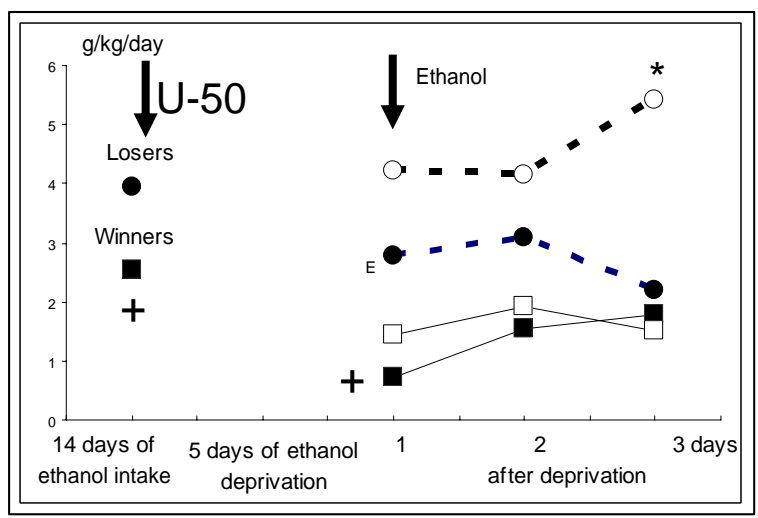

**Fig.7.** Effect of selective *k*-opioid receptors agonist U-50,488H (2.5 mg/kg) on the ethanol intake by the winners and losers [51]. Winners and losers were provided 10% ethanol under the free two bottles choice paradigm. By the end of a two-week period the losers started consuming significantly more ethanol as compared with the winners. During ethanol deprivation and after it each group of animals was treated with the drug (white marker) and vehicle (black markers). The drug produced no effect on ethanol intake in the winners. Losers were shown to increase the intake of ethanol.
*Note*: + p<0.05 – as compared with losers; * p<0.05 – as compared with vehicle treatment.

It was natural to presume that psychoemotional disorders induced by repeated experience of aggression and accompanied by victories (positive emotional background) or by repeated experience of social defeats (negative emotional background) would evoke different changes in the brain, triggering different reactions to the drugs as written above [8, 67]. Since the majority of the drugs used act on the receptors, it is natural to assume that receptors might become more sensitive to neurotransmitters (sensitization) and, thus, more susceptible to drugs, or less sensitive (desensitization). It is clear that such changes in the receptors arise in response to changes in mediator metabolism evoked by social confrontations. Therefore, on the one hand, psychoemotional states modify the effect of drugs, which is to be taken into consideration in the therapy. On the other hand, psychoemotional states themselves can exert significant influence on the development of a disease. For example, the growth of the Krebs-2 carcinoma and the metastasis of transplanted Lewis adenocarcinoma were significantly higher in the losers as compared with the winners and control (*Fig. 9*) [41, 43]. Apparent conclusion is that the development of new methods of rational cancer treatment should include a correction of the psychoemotional status of individuals, which may have an effect on both cancer growth and metastasis and the efficiency of cancer therapy.

It is also evident that different drugs may be needed to arrest the same modified behaviors or physiological changes in different individuals since such individuals may have varying susceptibility to those drugs. The proposed model allows investigating the responses of individuals with opposite psychoemotional states to the administration of the same drug. In modern pharmacology priority has been attached to personal therapy, which is thought to be a key to effective treatment.

### Conclusion: perspectives of the use of social models for pharmacological screening of psychotropic drugs

Over many years the studies have focused on finding adequate models of psychoemotional disorders [4, 13, 17-19, 30, 36, 38, 40, 44, 46, 74, 78, 80-82, 86, 94], which would allow screening of psychotropic drugs [16, 27, 29, 31, 78, 81, 86]. The models that are gaining in popularity include

*Table 5*. Individual sensitivity of the winners, losers and intact male mice to the effects of acute drug treatment.

| Drugs | Dose, treatment | Strain | Groups | Behavioral tests | Effects | References |
|---|---|---|---|---|---|---|
| **Flesinoxane,** 5-HT$_{1A}$ receptor agonist | 0.5 mg/kg, i.p., 30 min | C57 | Intact T10 winners T10 losers | Partition test* | Decrease Decrease No effect | [69] |
| **DAMGO,** mu-opioid receptor agonist | 2 mg/kg, s.c., 30 min 35 min | C57 | Intact | Partition test Plus-maze test | Decrease Anxiogenic | [53] |
| | | | T10 winners | Partition test Plus-maze test | Decrease Anxiogenic | |
| | | | T10 losers | Partition test Plus-maze test | No effect No effect | |
| **U-50,488H,** kappa-opioid receptor agonist | 0.6, 1.25, 2.5 mg/kg, s.c. 30 min 35 min | C57 | Intact | Partition test Plus-maze test | Decrease (1.25 mg/kg) Anxiolytic (2.5mg/kg) | [53] |
| | | | T10 winners | Partition test Plus-maze test | No effect No effect | |
| | | | T10 losers | Partition test Plus-maze test | Increase Anxiolytic (2.5mg/kg) | |
| **(+)3PPP,** D$_2$ receptor agonist | 2 mg/kg, i.p., 30 min | C57 | Intact T20 winners T20 losers | CPAR | No effect Inhibition Stimulation | [39] |

*Note:* CPAR – conditioned passive avoidance reaction. * Increase or decrease of behavioral reaction to the partner in the neighboring compartment in the partition test may involve different motivations in animals of different groups (control, winners and losers).

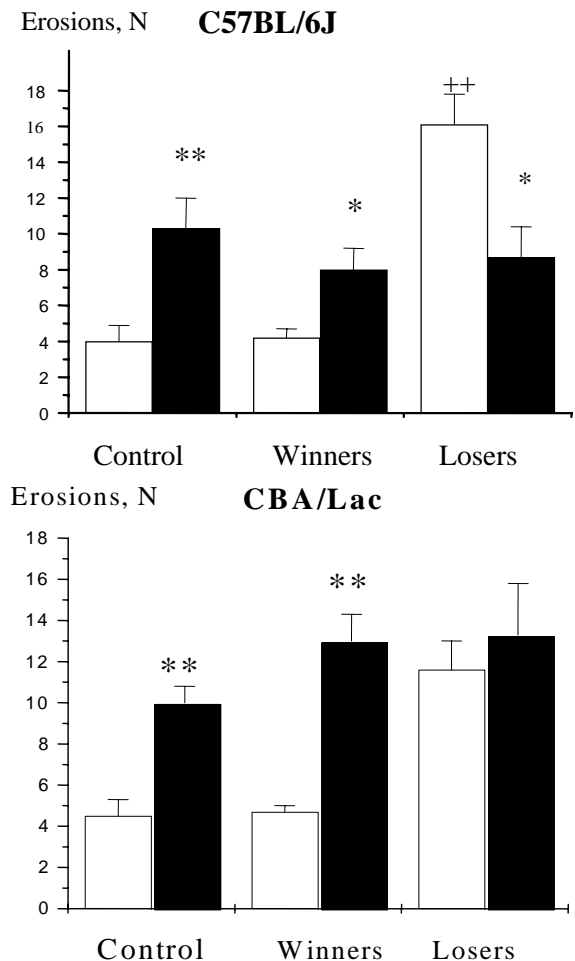

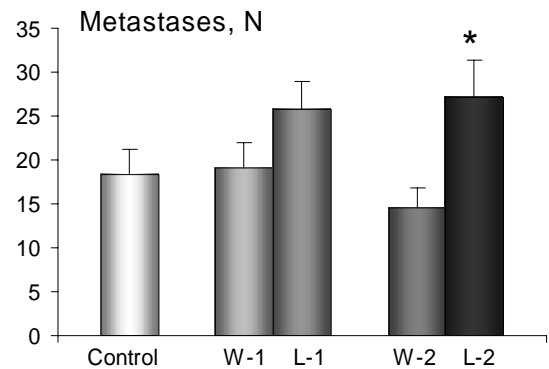

**Fig. 9.** Effect of psychoemotional status on the metastasis of Lewis carcinoma in mice [41]. W-1 and L-1 – winners and losers (respectively) who were injected with tumor cells after exposure to 20 days of confrontations; W-2 and L-2 – winners and losers (respectively) mice, which were injected with tumor cells after exposure to 6 days of agonistic interactions continued thereafter. The metastasis is shown to be affected by the social status: the losers had significantly more metastases in the lung as compared with the control and winners. * p<0,05 vs control and W-2

**Fig.8.** Effect of ciproheptadine administered into the 3d brain ventricle (10 mkg) on the development of gastric mucosa damage in C57 and CBA males after three-day social stress. Originally, the gastric mucosa damage in the losers of both strains was considerably higher than in the control and winners. Eighteen hours after drug injection the number of erosions increased in the control and winners of both strains, decreased in the C57 losers and did not change in the CBA losers. These data indicate that serotonin could produce a diverse effect in animals with different psychoemotional state [47].

*Note*: white columns – vehicle-treated animals; black columns – ciproheptadine-treated animals. * p<0.05; ** p<0.01 – as compared with vehicle; ++ p<0.01 – as compared with the control and winners (vehicle treated).

biosocial models to study the consequences of chronic social conflicts and social stress in animals [14, 15, 19, 28, 45, 78, 93]. In this connection our experimental behavioral approach is up to date and in the mainstream of contemporary studies. It allows screening of novel psychotropic drugs in simulated clinical conditions and detecting their preventive and therapeutic properties and efficacy. In our view, this method could help if not to entirely avoid but to minimize the phase of clinical trials of novel drugs on patients. A large variety of behavioral pathologies (anxious depression, catalepsy, reduced sociability, pronounced aggression, hyperkinesia, anxiety, hypersensitivity, hyperactivity, cognitive disturbances, anhedonia etc.), which are accompanied by somatic changes (gastric mucosa damage, reduced gonadal function, psychogenic immune deficiency and others) in animals gives grounds to suppose that this approach could be extensively used in many medical-biological studies.

Study of the state of the brain's mediator systems in each given moment of pathological process may also give valuable results. Experimental study of dynamic changes in brain neurotransmitter systems (metabolism, receptors, gene expression) – from norm to severe pathology could help to identify adequate methods of pharmacological correction and to suggest complex therapy depending on the stage of a disease. The action of drugs should be investigated with respect to the neurochemical background specifically altered under the influence of emotional pathogenic factors. Our studies on animals treated with psychotropic drugs that are commonly used in medical practice for treatment of depression and anxiety have shown a close correspondence of their effect to that in humans.

This thoroughly elaborated and highly productive sensory contact model with broad methodological capabilities is easy to implement. One of its strengths is relevance – similarity of psychogenic factors leading to the development of a disease, which are analogous to such factors in humans. Supposedly this model and pharmacological approach could be adopted by all pharmacological companies developing or screening novel psychotropic drugs with different spectra of action.